\documentclass[runningheads]{svmult}

\usepackage{makeidx}   
\usepackage{graphicx}  
\usepackage{subeqnar}  
\usepackage{multicol}  
\usepackage{physprbb}  
\makeindex             

\begin{document}
\title*{Probing Halos with PNe: 
Mass and Angular Momentum in Early-Type Galaxies}
\toctitle{Probing Halos with PNe: 
Mass and Angular Momentum in Early-Type Galaxies}
\titlerunning{Probing Halos with PNe}
\author{Aaron J. Romanowsky
}
\authorrunning{Aaron Romanowsky}
\institute{School of Physics and Astronomy, University of Nottingham,
University Park, Nottingham NG7 2RD, England
}

\maketitle              

\begin{abstract}
We present an observational survey program
using planetary nebulae, globular clusters, and
X-ray emission to probe the halos of early-type galaxies.
We review evidence for scanty dark matter halos around ordinary
elliptical galaxies,
and discuss the possible implications.
We also present measurements of rotation in the halos.
\end{abstract}

\section{Introduction}
Studying early-type galaxies (ellipticals and lenticulars)
in their outer parts (much outside $\sim R_{\rm eff}$, the half-light radius)
can provide powerful clues about their structure and formation -- but
the very low stellar surface densities and the general lack of cold gas
have precluded a convenient observational approach, and progress in this area has been slow.
However, the situation is now changing with recent advances in instrumentation (described below) 
and the use of planetary nebulae (PNe), globular clusters (GCs), and X-ray emission as halo probes.

As discussed in Sect. 3, some initial results are puzzling:
PN and GC dynamics in ordinary $L^*$ ellipticals imply much less dark matter (DM) 
than expected in theories of galaxy formation.
However, other lines of observational evidence -- using
satellite dynamics and weak lensing \cite{sat} -- imply more dominant DM halos
in $L^*$ ellipticals.
Also, spiral galaxies and very bright ellipticals show massive DM halos \cite{spiral}.
Thus, not only is it necessary to study a larger sample of galaxies
for better statistics,
but it is also important to cross-check the different mass probes for consistency.
After identifying which probes are reliable, we can combine them for much stronger
constraints on the mass and orbit distributions in galaxy halos.
To explore these issues, we are studying a representative sample of near-$L^*$ ellipticals
using integrated stellar kinematics, PNe, GCs, and X-rays.

\section{Halo Probes: PNe, GCs, and X-rays}

\begin{figure}[t]
\begin{center}
\includegraphics[width=.496\textwidth]{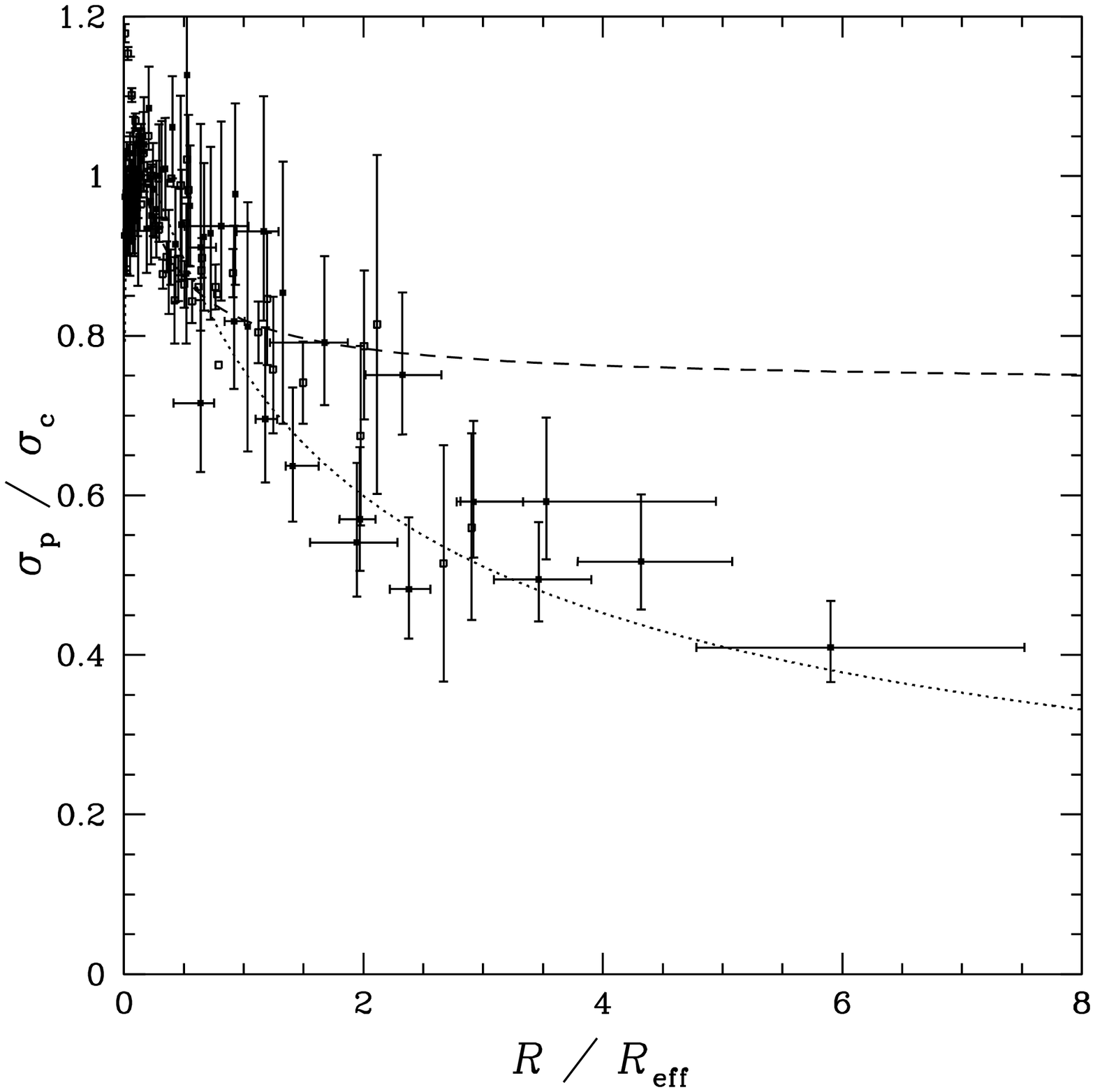}
\includegraphics[width=.496\textwidth]{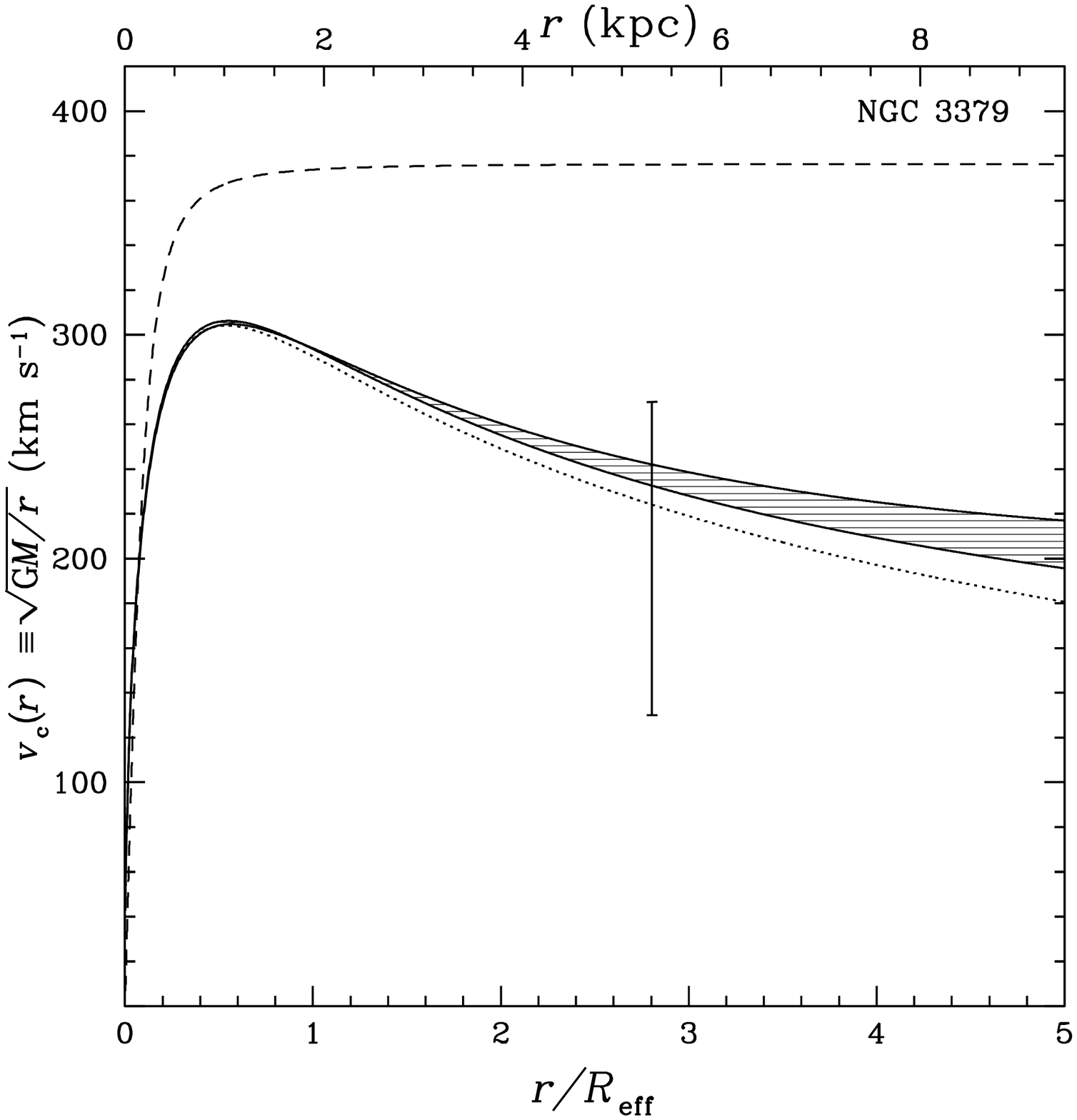}
\end{center}
\caption[]{
{\it Left:}
Projected velocity dispersion profiles with radius for four elliptical galaxies 
\cite{mend,roman03},
scaled and stacked.
Simple model predictions show an isothermal halo (dashed line)
and a constant mass-to-light ratio (${\rm \Upsilon}$) galaxy (dotted line).
{\it Right:}
Modeled circular velocity profile of NGC~3379.
Shown are a constant ${\rm \Upsilon}$ solution (dotted line);
results from PN dynamics (shaded region \cite{roman03}),
from an X-ray analysis (dashed line \cite{osull04b}),
and from GC dynamics assuming orbital isotropy (error bar \cite{n3379gcs}).
}
\label{comb1}
\end{figure}

{\bf PN} kinematics are a very promising probe, tracing the main stellar population
in galaxy halos which is otherwise too diffuse to observe.
The {\it PN.Spectrograph} is the breakthrough instrument for this technique,
making it now possible to obtain hundreds of PN velocities in $L^*$ galaxies out to
distances of $\sim$~25 Mpc \cite{doug}.
Four $L^*$ ellipticals have been studied using PNe, and their
projected velocity dispersions have been found to decline markedly with galactocentric radius
(see Fig.~\ref{comb1}, {\it left}), suggesting very low DM content.

Like PNe, {\bf GCs} are useful as bright point-like sources in galaxy halos.
GC systems are typically more extended than the main galaxy light,
allowing them to probe to even larger radii than PNe.
Furthermore, GCs are interesting in their own right as simple stellar systems
formed at early times:
their properties can reveal events in their host galaxy's history which
are difficult to discern in the jumble of the main galaxy light.
One generic prediction of most GC formation scenarios is that the more
metal-rich GCs are associated with the main stellar population -- which can be
verified by comparisons of metallicities and kinematics between the
halo stars (represented by the PNe) and the GCs.

While it has long been possible to study GC kinematics in very GC-rich galaxies like
M87, it is only with the latest spectrographs that this is viable for
a large sample of typical galaxies.  These instruments
(e.g., VLT+FLAMES,VIMOS; Magellan+IMACS; Gemini+GMOS; Keck+DEIMOS)
have the advantages of a wide field of view, multi-object capability, and
8-meter-class collecting area.

In principle, thermal {\bf X-ray emission} from hot gas trapped in the potential well of
a galaxy offers an excellent probe of its halo mass, 
as it is not subject to the systematic anisotropy uncertainty plaguing stellar systems.
In practice, it is difficult and crucial to remove contaminating point sources, to
check the dynamical equilibrium of the gas,
and to measure the radial temperature variation.
With their large collecting areas and high angular resolution, the {\it Chandra}
and {\it XMM-Newton} observatories have made it now possible to rigorously pursue mass studies based
on X-rays.
We note the importance of not selecting the galaxy sample based on $L_{\rm X}$
as this will introduce a bias in the halo mass results.

\section{Mass Results}

We have collected published results on the mass profiles of
various early-type galaxies (M87, NGCs 1399, 3379, 4472, 4636, 5128)
to check their consistency; the techniques include PNe, GCs, and X-rays 
\cite{gal}.
The results are generally consistent, even in cases where parts of the halo
appear to be well out of equilibrium.

A case study is NGC~3379,
for which
we have constructed dynamical models to fit long-slit stellar kinematical
data in combination with PN velocities \cite{roman03};
the models fully include the variations in orbital anisotropy which otherwise
make the mass profile very uncertain.
We find a total $B$-band mass-to-light ratio at 5~$R_{\rm eff}$ of ${\rm \Upsilon}_B=7.1\pm0.6$ (solar units);
for comparison, standard models of stellar populations
imply ${\rm \Upsilon}_{*,B}=$~6--11,
leaving very little room for DM inside 5~$R_{\rm eff}$.

A major remaining systematic uncertainty in this analysis is the assumption that
NGC~3379 is approximately spherical.  But if it
in fact contained a significant disk component (as in an S0 galaxy) viewed face-on,
the large-radius stellar kinematics could appear ``cold''
even with a massive DM halo.
Although the galaxy's inner parts do not show the disk-like kinematics expected in
such a scenario \cite{statler},
a more convincing test is to use independent mass probes
(Fig.~\ref{comb1}, {\it right}).
The GCs are a good candidate since GC systems are rarely flattened;
GC kinematics in NGC 3379
reveal a low halo velocity dispersion consistent with the PN result \cite{n3379gcs}.
A ring of HI gas at large radius also shows kinematics consistent with the
low DM content \cite{schneider}.
On the other hand, a preliminary analysis of the galaxy's X-ray emission
using {\it Chandra} data
indicates a higher DM content \cite{osull04b}.
We will see if these various analyses,
when complete, converge to similar results.

NGC~3379 seems a convincing case for a low-DM system, but
how common is this situation?
Another candidate is the S0 galaxy NGC~5128, where PN and GC dynamics
imply a total ${\rm \Upsilon}_B\sim$~20 inside the virial radius 
\cite{peng04}.
But it and the other three $L^*$ ellipticals still need to be modeled in as much
detail as NGC~3379 to verify the apparent low DM content.
Note also that one of these, NGC~4494, has very faint gaseous
X-ray emission, consistent with the low-DM scenario \cite{osull04a}.

Firm conclusions await a larger sample size along with rigorous
analyses, but if the apparent low DM content of these galaxies turns out to be
a widespread phenomenon, what could be the explanation?
The idea that these are all face-on S0s is unlikely from a statistical
point of view and because of the cross-checks in NGC~3379.
The galaxies may have lost much of their DM halos through interactions
with other galaxies, but most of them are not in the high-density
environments thought necessary for this.
Modified Newtonian Dynamics has been advanced as an explanation \cite{milgrom},
but this alternative gravitational theory has 
difficulty explaining the large mass discrepancies found
in brighter ellipticals.

A higher DM content in these galaxies may be possible if
baryonic processes have driven more DM into the galaxy centers than
we have assumed using standard ${\rm \Lambda}$CDM halos \cite{nfw}.
But this would contradict many other lines of evidence that
the DM mass fraction in the inner parts of ellipticals is fairly low,
and would involve lower values for ${\rm \Upsilon}_*$ than
are plausible using current stellar population synthesis models.
An additional possibility is that a decline with radius
of ${\rm \Upsilon_*}$ could mask the increase from a DM halo;
however, this would require either a
metallicity log gradient of $\sim -1.4$ -- which
is much larger than what is implied by typical studies of integrated stellar light 
 and by the abundance studies of halo PNe presented at this workshop -- or
a young halo population, which is excluded in NGC~3379
by study of the color-magnitude diagram of resolved stars \cite{gregg}.

A more arguable scenario is that these galaxies do in fact have large
amounts of DM residing even farther out in the halo than we have been
able to probe.  Such halos would have much lower concentrations 
(or equivalently, lower central DM mass fractions)
than predicted by ${\rm \Lambda}$CDM.
This conclusion has also (controversially) been reached with
many dynamical studies of late-type galaxies 
and with strong lensing studies of early-types,
and could be explained by new baryonic processes 
or by yet more exotic DM theories \cite{lowc}.
We are now quantifying constraints on DM halo models 
vis-\`{a}-vis empirical results on ${\rm \Upsilon}$ gradients \cite{nap04}.

\section{Angular Momentum Results}

It is well known that elliptical galaxies exhibit much smaller
specific angular momenta $\lambda$ than spirals.
But studies in ellipticals have been confined to their central
parts, and it is surmised that the ``missing'' $\lambda$ may
be found in their outskirts.
Indeed, most simulations of elliptical formation through
galaxy major mergers predict
that their outer parts should rotate rapidly.

\begin{figure}[t]
\begin{center}
\includegraphics[width=.496\textwidth]{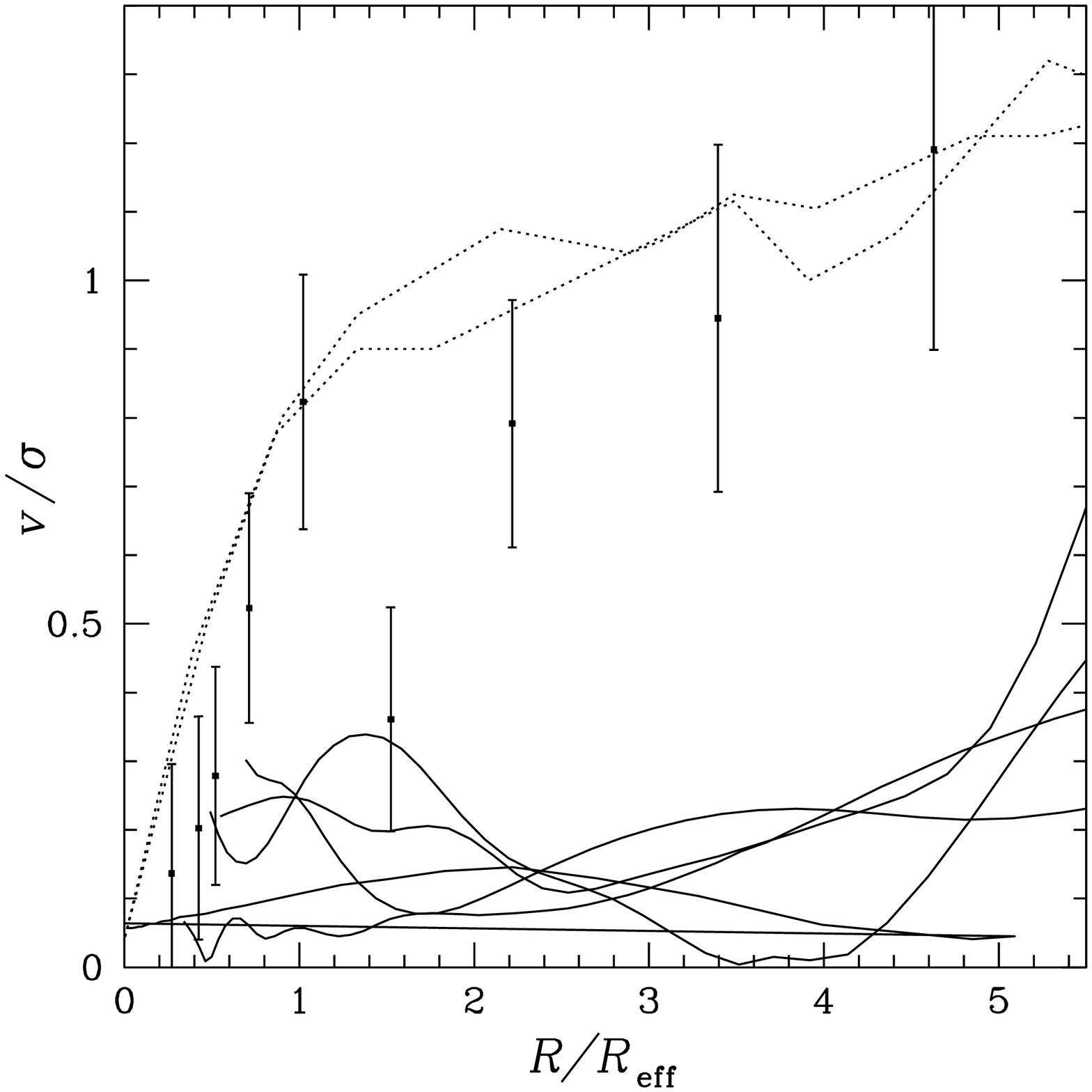}
\includegraphics[width=.496\textwidth]{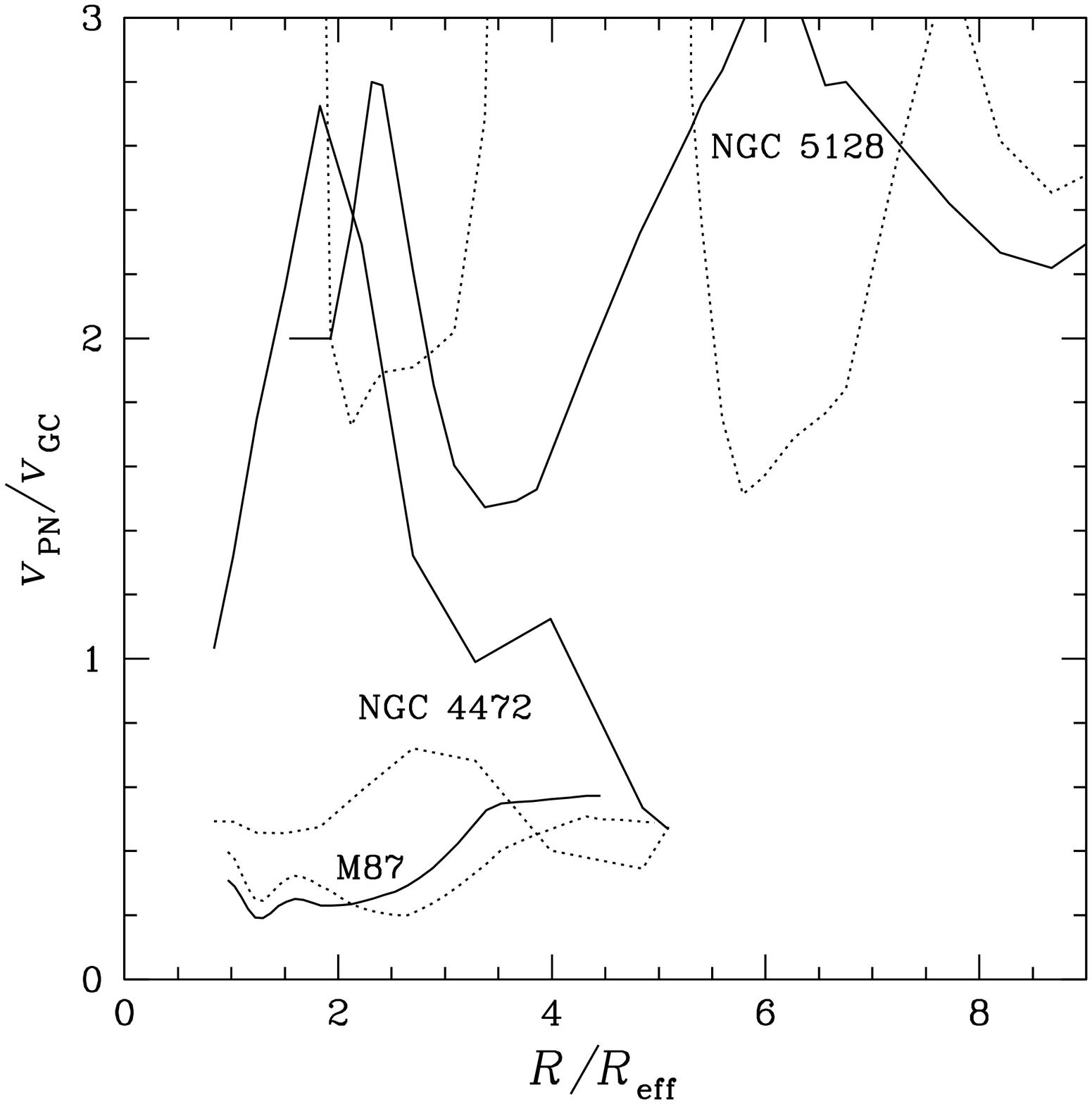}
\end{center}
\caption[]{
{\it Left:} Rotational parameter with radius for five elliptical galaxies 
(M87, NGC 821, NGC 3379, NGC 4472, and NGC 4494)
observed with PNe (solid lines \cite{roman03,roman04}), 
for NGC 5128 (error bars \cite{peng04}),
and for a simulated merger remnant (dotted lines \cite{weil}).
{\it Right:}
Ratio between PN and GC rotational velocities at a given radius,
in three galaxies.
Dotted lines are for metal-poor GCs, and solid lines for metal-rich GCs.
}
\label{vsig}
\end{figure}

With PN data in five ellipticals, we can now examine
the rotation of their stellar halos (Fig.~\ref{vsig}, {\it left}).
In none of these galaxies do we find substantial rotation:
typically, $v/\sigma\sim$~0.2.
On the other hand, NGC~5128 has a rapidly rotating halo ($v/\sigma\sim1$),
consistent with a dissipationless merger model \cite{weil}.
A caveat is that the ellipticals have been
partially selected for their roundness, biasing against
observably high rotation; 
more secure results will come with corrections for projection effects,
and with a wider ellipticity range in the galaxy survey.
Also, the theoretical expectations are not yet entirely clear;
inclusion of baryonic physics seems to produce galaxies with
lower $\lambda$ than in dissipationless simulations \cite{sims}.

We can also compare the rotational behavior of the stars (via the PNe)
with the GCs (Fig.~\ref{vsig}, {\it right}).
However, with the few cases so far
available for study, we do not yet see a consistent pattern
(such as $v_{\rm PN}/v_{\rm GC}\sim$~1 for metal-rich GCs).

\section{Summary}

Comparisons of early-type galaxy halo mass results from independent probes
(PNe, GCs, and X-rays) yield generally consistent results.
A surprising trend is seen for $L^*$ ellipticals to show
low dark matter content, which may be indicative of
low-concentration halos.
We find so far little evidence for the strong rotation 
that may be expected in ellipticals' stellar halos.

%


\begin{thebibliography}{8.}
\addcontentsline{toc}{section}{References}

\bibitem{sat} F. Prada et al.: ApJ \textbf{598}, 260 (2003); J. Guzik, U. Seljak: MNRAS \textbf{335}, 311 (2002)

\bibitem{spiral} M. Persic, P. Salucci, F. Stel: MNRAS \textbf{281}, 27 (1996); M. Loewenstein, R. White: ApJ \textbf{518}, 50 (1999); T. Treu, L. Koopmans: ApJ {\bf 611}, 739 (2004)

\bibitem{doug} N. Douglas: this volume; N. Douglas et al.: PASP \textbf{114}, 1234 (2002)

\bibitem{mend} R. M\'{e}ndez, A. Riffeser, R.-P. Kudritzki, M. Matthias et al.: ApJ \textbf{563}, 135 (2001)

\bibitem{roman03} A. Romanowsky, N. Douglas, M. Arnaboldi et al.: Science \textbf{301}, 1696 (2003)

\bibitem{osull04b} E. O'Sullivan, A. Romanowsky, T. Ponman et al.: in preparation

\bibitem{n3379gcs} M. Beasley et al.: in prep; G. Bergond, A. Romanowsky, S. Zepf et al.: in prep

\bibitem{gal} A. Romanowsky, C. Kochanek: ApJ: 553, 722 (2001); K. Matsushita et al.: A\&A \textbf{386}, 77 (2002); 
Y. Ikebe et al.: Nature \textbf{379}, 427 (1996); C. Jones et al.: ApJ \textbf{482},~143 (1997); M. Paolillo et al.: ApJ \textbf{565}, 883 (2002); N. Napolitano et al.: A\&A \textbf{383},~791 (2002); T. Richtler et al.: AJ \textbf{127}, 2094 (2004);
S. Schindler et al.: A\&A \textbf{343}, 420 (1999);
A. Kronawitter et al.: A\&AS \textbf{144}, 53 (2000);
K. Matsushita et al.: ApJ \textbf{499}, L13 (1998); M. Loewenstein, R. Mushotzky: 
{\tt astro-ph/0208090}; Y.~Schuberth et al.: this volume

\bibitem{statler} T. Statler: AJ \textbf{121}, 244 (2001)

\bibitem{schneider} S. Schneider: ApJ \textbf{288}, L33 (1985)

\bibitem{peng04} E. Peng, H. Ford, K. Freeman: ApJ \textbf{602}, 685 (2004); ApJ \textbf{602}, 705 (2004)

\bibitem{osull04a} E. O'Sullivan, T. Ponman: MNRAS \textbf{349}, 535 (2004)

\bibitem{milgrom} M. Milgrom, R. Sanders: ApJ \textbf{599}, L25 (2003)

\bibitem{nfw} J. Navarro, C. Frenk, S. White: ApJ \textbf{490}, 493 (1997)

\bibitem{gregg} M. Gregg, H. Ferguson, D. Minniti, N. Tanvir, R. Catchpole: AJ \textbf{127}, 1441 (2004)

\bibitem{lowc} J. Ostriker, P. Steinhardt: Science \textbf{300}, 1909 (2003); D. Rusin et al.: ApJ \textbf{595}, 29 (2003); H. Mo, S. Mao: MNRAS {\bf 353}, 829 (2004)

\bibitem{nap04} N. Napolitano et al.: this volume; MNRAS, in press ({\tt astro-ph/0411639})

\bibitem{roman04} A. Romanowsky, N. Douglas, K. Kuijken, M. Arnaboldi: in preparation

\bibitem{weil} M. Weil, L. Hernquist: ApJ \textbf{460}, 101 (1996)

\bibitem{sims} J. Sommer-Larsen et al.: ApJ \textbf{596}, 47 (2003);
L. Wright et al.: {\tt astro-ph/0310513}


\end{thebibliography}
\end{document}